\documentclass[9pt,conference]{IEEEtran}
\usepackage{cite}
\usepackage{url}
\usepackage{amsmath,amssymb,amsfonts}
\usepackage{algorithmic}
\usepackage{graphicx}
\usepackage{textcomp}
\usepackage{boldline}
\usepackage{xcolor}
\def\BibTeX{{\rm B\kern-.05em{\sc i\kern-.025em b}\kern-.08em
    T\kern-.1667em\lower.7ex\hbox{E}\kern-.125emX}}
\vspace{-0.2in}
\begin{document}

\title{POSTER: Diagnosis of COVID-19 through Transfer Learning Techniques on CT Scans: A Comparison of Deep Learning Models\\}

\author{\IEEEauthorblockN{Aeyan Ashraf}
\IEEEauthorblockA{\textit{Department of Computer Engineering} \\
\textit{Aligarh Muslim University}\\
Aligarh, India \\
gj3323@myamu.ac.in}
\and
\IEEEauthorblockN{Asad Malik}
\IEEEauthorblockA{\textit{Department of Computer Science} \\
\textit{Aligarh Muslim University}\\
Aligarh, India \\
amalik\_co@myamu.ac.in}
\and
\IEEEauthorblockN{Zahid Khan}
\IEEEauthorblockA{\textit{College of Computer and Information Sciences} \\
\textit{Prince Sultan University}\\
Riyadh, Saudi Arabia\\
zskhan@psu.edu.sa}
}
\maketitle
\begin{abstract}
The novel coronavirus disease (COVID-19) constitutes a public health emergency globally. It is a deadly disease which has infected more than 230 million people worldwide. Therefore, early and unswerving detection of COVID-19 is necessary. Evidence of this virus is most commonly being tested by RT-PCR test. This test is not 100\% reliable as it is known to give false positives and false negatives. Other methods like X-Ray images or CT scans show the detailed imaging of lungs and have been proven more reliable. This paper compares different deep learning models used to detect COVID-19 through transfer learning technique on CT scan dataset. VGG-16 outperforms all the other models achieving an accuracy of 85.33\% on the dataset.
\end{abstract}
\section{Introduction}
In 2020, Severe Acute Respiratory Syndrome Coronavirus 2 abbreviated as SARS-CoV-2, an extremely dangerous and previously unknown virus, wreaked havoc across the globe, killing more than 4.5 million people and leaving life long complications among many. It was declared a pandemic by WHO on March 11, 2020\cite{wang2020novel}. COVID-19 is a disease caused by the SARS-CoV-2 virus. Coronavirus's first case was found in Wuhan, China, in December 2019. The RT-PCR (Real-Time Polymerase Chain Reaction) test used to test the evidence for this virus is not 100\% reliable as it is known to give false-negative and false-positive results. Furthermore, it takes at least 6-8 hours to give the results, which is not helpful as this virus spreads rapidly through the air. In comparison to RT-PCR, detection via CT scans is more reliable, useful, and quicker technology for the classification of COVID-19. Recently, computer vision and image processing has led to extraordinary developments with the advancement of deep learning and especially the development of convolutional neural networks (CNN). The AI-based computer-aided system can speed up the diagnostic process and identify details imperceptible to the human eye.
Several studies have been done in this domain. Meng Li\cite{li2020chest} reviewed the chest CT features of COVID-19 patients and analyzed the role of Chest CT. Dai et al.\cite{dai2020ct} explained the findings from lung CT scans of some patients that were admitted in The Second People's Hospital of Shenzhen for treatment of COVID-19. Fang et al.\cite{fang2020sensitivity} found that the chest CT scans demonstrated a higher sensitivity of 97\% and was therefore a better way of screening and monitoring by radiologists. Ai et al.\cite{ai2020correlation} studied the correlation of Chest CT and RT-PCR Testing for Coronavirus disease. Nuriel\cite{mor2020corona} also constructed a CNN model based on MobileNetV2 to evaluate the ability of deep learning to detect COVID-19 from chest CT images. Yu et al.\cite{chen2020deep} constructed a system based on UNet++ for identification of COVID-19 from CT images. Chand et al.\cite{10.1007/978-981-16-2919-8_35} compared AlexNet, Resnet, VGG for classification of COVID-19. Similarly research has been carried out for detecting COVID-19 from X-Ray images as well(\cite{jawahar2021utilization} and \cite{10.1007/978-981-16-3690-5_37} are some examples).
\section{Methods and Materials}
\subsection{Dataset used}
This paper uses an open-sourced dataset COVID-CT\cite{zhao2020COVID-CT-Dataset}, which contains 349 COVID-19 CT images from 216 patients and 463 non-COVID-19 CTs. The utility of this dataset has been confirmed by a senior radiologist who was diagnosing and treating COVID-19 patients since the outbreak of the pandemic.
\subsection{Data Preprocessing}
%\subsubsection{Resizing}
Since the images were of different sizes, we resized them to $224 \times 224$. For this we used the Keras ``preprocess\_input'' function. The dataset was not very large so to address the problem of overfitting various data augmentation techniques were applied to the training samples. ``ImageDataGenerator'' function was used for this purpose. We set the rotation range to 20 degrees and used Random Height, Random Width and Random Flip.
%\begin{figure}[http!]
%	   \centering
%	   \includegraphics[scale=0.3]{fig/yo.png}
%       \caption{An example of Augmentation applied by fixing rotation range to 20 degrees: Random Height (a)-(c), Random Width (d)-(f), Random Flip (g)-(i).}
%	   \label{fig:1}
%\end{figure}
%\subsubsection{Augmentation}
%\subsection{Deep Learning}
%Deep Learning is the sub branch of Machine Learning in which machine imitates the way humans gain knowledge. It uses multiple layers to progressively extract high-level features from raw input.
%\subsection{Convolutional neural network}
%Convolutional neural network abbreviated as CNN is a type of Neural network which is used to process or analyze visual data like images, video etc. They are used in image and video recognition, image classification, segmentation, medical image analysis, Natural Language processing etc.
\subsection{Transfer learning}
\begin{figure}[http!]
	   \centering
	   \includegraphics[scale=0.22]{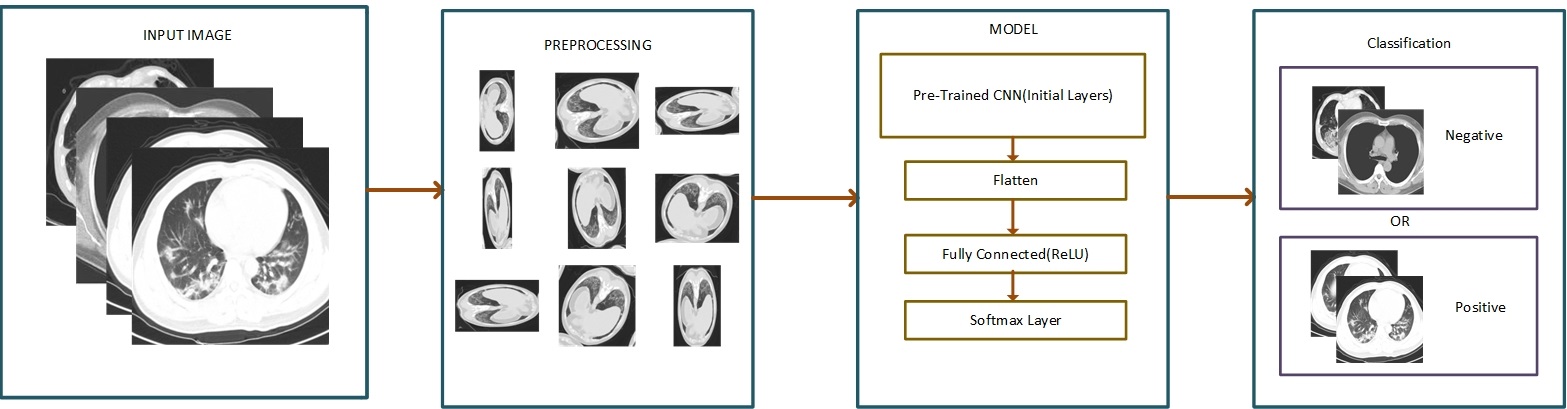}
        \caption{A basic work flow of transfer learning model by using augmentation technique.}
	   \label{fig:1}
\end{figure}

Transfer Learning is a method which helps us in building an accurate model without starting the process from scratch. It saves us a lot of time and is useful because we use pre-trained models which does not need a large dataset. In this paper we have used VGG16 \& VGG19\cite{simonyan2014very}, Resnet50\cite{he2016deep}, InceptionV3\cite{szegedy2016rethinking}, Xception\cite{chollet2017xception} models. The input size for all the models is (224 $\times$ 224 $\times$ 3). The problem that we usually face in using a pre-trained model is overfitting. In our case we used data Augmentation and Dropout to address this issue. After the convolutional base the flatten layer is placed which transforms the 2-D feature matrix into a vector. After that fully connected layer with Rectified Linear Unit(ReLU) is used. The output is then fed to Softmax activation Layer for final classification. Fig.~\ref{fig:1}   shows how a model is constructed using transfer learning.
\section{Experiments and Results }
For this experiment we have used the Google Colaboratory also known as Colab with Intel(R) Xeon(R) CPU @2.30GHz, 30GB RAM and NVIDIA TESLA P100 GPU. The performance evaluation used to see the performance of the model in this research was calculated using AUC(area under the receiver operating characteristic(ROC) curve), Accuracy, Precision, Recall, Specificity, $F_{1}$ score, Computation Time. %These metrics are calculated using the following equations:
%\begin{equation}Accuracy=\frac{TP + TN}{TP+FP+TF+FN} \times 100\% \end{equation}
%\begin{equation}Precision=\frac{TP}{TP+FP} \times 100\% \end{equation}
%\begin{equation}Recall=\frac{TP}{TP+FN} \times 100\% \end{equation}
%\begin{equation}Specificity=\frac{TN}{TN+FP} \times 100\% \end{equation}
%\begin{equation}F_{1}=2\times\frac{Precision\times Recall}{Precision + Recall} \end{equation}
%where TP: True Positive; TN: True Negative; FN: False Negative, and FF: False Positive.
%\begin{figure}[http!]
%	   \centering
%	   \includegraphics[width=9cm,height=4cm]{fig/accuracy.jpg}
%        \caption{Accuracy}
%	   \label{fig:3}
%\end{figure}
%\begin{figure}[http]
%	   \centering
%	   \includegraphics[width=9cm,height=4cm]{fig/precision.jpg}
  %      \caption{Precision}
%	   \label{fig:6}
%end{figure}
%\begin{figure}[http]
%	   \centering
%	   \includegraphics[width=9cm,height=4cm]{fig/recall.jpg}
%        \caption{Recall}
%	   \label{fig:7}
%\end{figure}
%\begin{figure}[http!]
%	   \centering
%	   \includegraphics[width=9cm,height=4cm]{fig/f1.jpg}
%        \caption{$F_{1}$ Score}
%	   \label{fig:4}
%\end{figure}
%\begin{figure}[http]
%	   \includegraphics[width=9cm,height=4cm]{fig/specificity.jpg}
%      \caption{Specificity}
%	   \label{fig:9}
%\end{figure}
%\begin{figure}[http]
%	   \centering
%	   \includegraphics[width=9cm,height=4cm]{fig/auc.jpg}
%        \caption{Area under the ROC Curve}
%	   \label{fig:10}
%\end{figure}
\begin{table}[http!]
	\caption{Model performance with different epochs.}
	\centering
\resizebox{1\columnwidth}{!}{
		\begin{tabular}{|l| l| l| l| l| l| l| }
        \hlineB{2}
            \bf Model 	 & \bf 50-epochs& \bf 75-epochs &   \bf 100-epochs &\bf best-ACC&\bf best-F1&\bf best-AUC\\
          \hlineB{2}
            VGG-16       &      85.33\%      &     82\%	  &   81.33\%        &  85.33\%    &  85.51\%    &	0.858	\\	
            \hline
            VGG-19       &      82\%      &     84.66\%	  &   81.33\%        &  84.66\%    &  83.68\%    &	0.846	\\	
            \hline
             ResNet-50       &      49.33\%      &     51.33\%	  &   63.33\%        &  63.33\%    &  69.6\%    &	0.65	\\	
            \hline
            InceptionV3       &      76\%      &     76.66\%	  &   76\%        &  76.66\%    &  76.62\%    &	0.765	\\	
            \hline
            Xception       &      78.66\%      &     78.66\%	  &   78.66\%        &  78.66\%    &  80\%    &	0.794	\\	
           \hlineB{2}
		\end{tabular}}
	\label{table:1}
\end{table}
\begin{figure}[http!]
   \centering
	   \includegraphics[height=6cm, width=9cm]{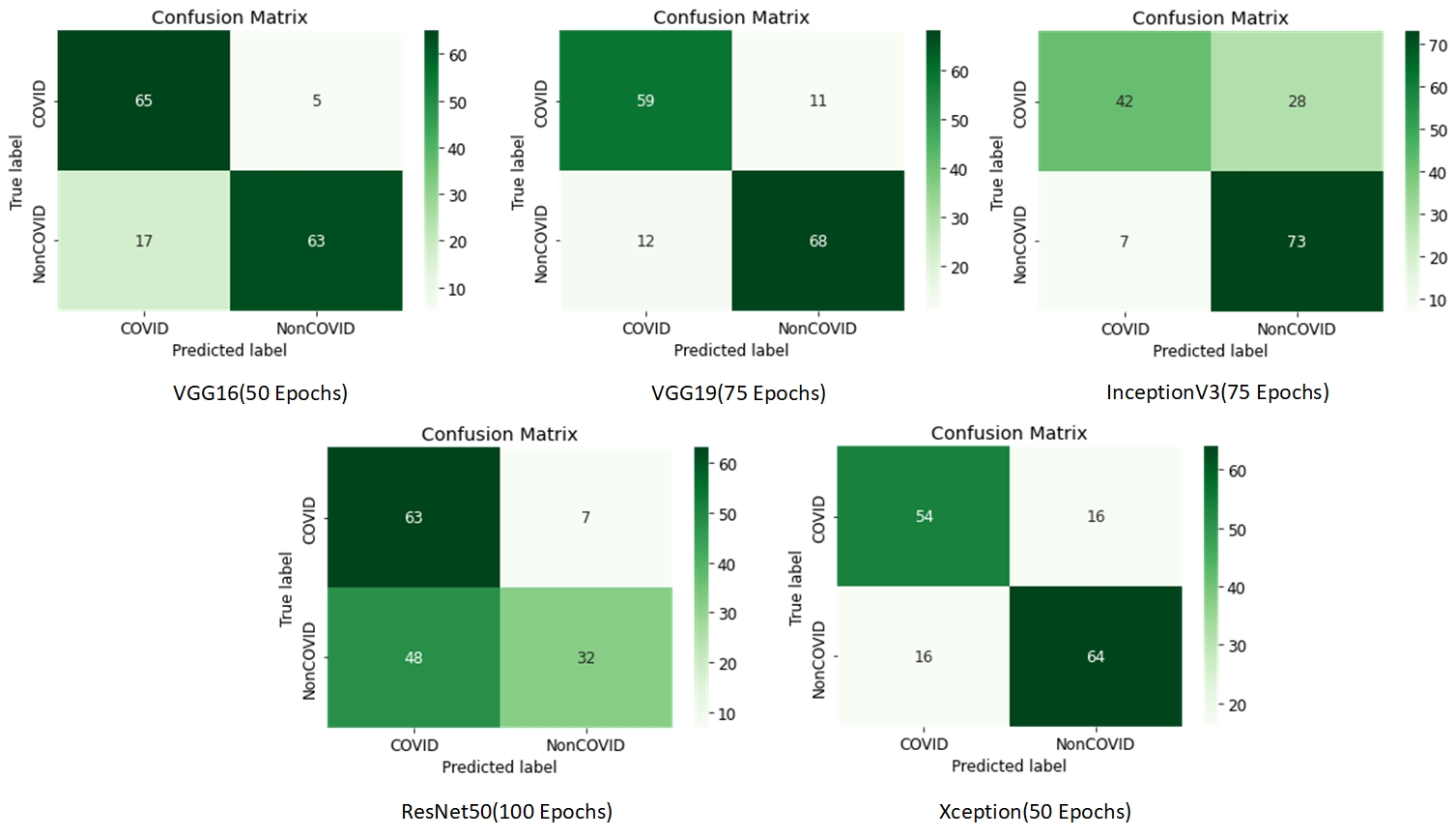}
        \caption{Confusion Matrices}
	   \label{fig:2}
\end{figure}
Table~\ref{table:1} shows the accuracy, $F_{1}$ score and AUC for each model. Among all the tested models, VGG16 and VGG19 produced the best classification accuracy with 85.33\% and 84.66\% respectively, whereas ResNet50 produced the lowest with 63.33\%. InceptionV3 and Xception showed considerable performance with 76.66\% and 78.66\%. It can be observed that the highest $F_{1}$ score has been attained by VGG16, which is 85.51\% followed by VGG19 with 83.68\%. The lowest $F_{1}$ score was attained by ResNet50 with 64.47\%, followed by InceptionV3 with 70.58\% and Xception with 80\%.

The confusion matrices(in their respective best cases) for the models are shown in \ref{fig:2}. If we take a look at the confusion matrices, VGG16 has the most correct classifications followed by VGG19. Resnet50 classified the COVID images very well but performed poorly on NonCOVID images. InceptionV3 performs well on the NonCOVID images but can classify very few COVID images correctly. Xception performs fairly. If Computation time is considered InceptionV3 performs the best with 306ms/step followed by VGG16(310ms/step), VGG19(315ms/step), Xception(345ms/step) and lastly ResNet50(360ms/step). Since InceptionV3 has inferior accuracy \& is only faster by a very small margin(5ms/step) therefore VGG16 and VGG19 prove to be the best models for diagnosing COVID-19 from CT scans.
\section{Conclusion}
 In this experiment we trained 5 different models with variations in epochs(50, 75 and 100). There are a lot of factors which influence the performance of the models like image quantity, Model complexity, optimizer, epochs, loss function. Our experiments and result obtained shows that VGG models perform the best for classification on this particular dataset. ResNet50 has a simple, single-scale processing unit with data pass-through connections due to which it takes a toll on it's performance. InceptionV3 divides processing by scale, merges the results, and repeats leading to the gain in performance versus Resnet. Xception and InceptionV3 are similar as Xception stands for extreme Inception. It uses depthwise separable convolutions. From this we can conclude that the very shallow and simple models like VGG perform well on our dataset. Models with shortcut connection adding the input of the block to its output like ResNet50 perform very poorly. Wider models like InceptionV3 and Xception perform fairly good.
\bibliographystyle{IEEEtran}

\small{
    \bibliography{reference}
}

\end{document}